# La$_2$Sb, a layered superconductor with metal–metal bonds

Hiroshi Mizoguchi and Hideo Hosono

Frontier Research Center, Tokyo Institute of Technology, 4259 Nagatsuta, Midori-ku, Yokohama 226-8503, Japan

Corresponding author: Hideo Hosono

E-mail: hosono@lucid.msl.titech.ac.jp

**Abstract**

**We found La$_2$Sb with a layered structure composed of alternate stacking of La square nets and LaSb layers exhibits bulk superconductivity with a critical temperature of 5.3 K. This suggests that the presence of the square net with strong La–La metal bonding is essential for the emergence of superconductivity.**

Since the discovery of high temperature superconductivity in F-doped LaFeAsO,[1] iron-based layered pnictides including $AFe_2As_2$ (A=Sr, Ba), and AFeAs (A=Li) has attracted much attention from solid state physicists and chemists. Extensive efforts have also been devoted to increasing the transition temperature $T_c$.[2] This has led to the rapid rise of $T_c$ to 56 K in $SmFeAs(O_{1-x}F_x)$,[3] the highest $T_c$ other than those for cuprate superconductors. These pnidtides or chalcogenides adopt a tetragonal structure that includes the FeAs layer. Therefore, the physical properties are expected to be two dimensional (2D) in nature, similar to other superconductors such as cuprate, $Sr_2RuO_4$, and $Na_xCoO_2$ $yH_2O$.[4] We have recently focused on layered metal pnictides because the relatively small electronegativity difference between the metal and pnictgen leads to wide energy bands near Fermi energy ($E_F$) to form a 2D metallic state. Therefore, we expect to have a possible platform for the exploration of new superconductors; even though $T_c$ is only moderately high in this system. It will provide complementary information to better understand the mechanisms of superconductivity and give clues for exploring new material systems for higher $T_c$ superconductors. In this communication, we report a new layered pnictide superconductor, $La_2Sb$, which is a metal-rich pnictides containing several kinds of metal–metal bonds. Although the synthesis and structure of this compound have been known since 1970,[5] the electrical and magnetic properties have not been previously reported.

Polycrystalline samples of $La_2Sb$ were synthesized using solid state reactions at elevated temperatures in evacuated silica ampoules. The starting materials used were La (99.9%), and Sb (99.9%). Stoichiometric amounts of these mixtures were heated in an evacuated silica ampoule at 773 K for 10 h, followed by heat-treatment at 1073 K for 10 h. The products obtained were ground and pressed into pellets. This was followed by additional heat-treatment in an evacuated silica ampoule at 1173 K for 20 h. All the treatments of the starting materials were performed in an Ar-filled glove box ($O_2$, $H_2O$ < 1 ppm). The crystal structure of the synthesized materials was examined using powder X-ray diffraction (XRD; Bruker D8 Advance TXS, Karlsruhe, Germany) using Cu K$_\alpha$ radiation with the aid of RIETVELD refinement using Code TOPAS3.[6] The temperature dependence of the DC electrical resistivity ($\rho$) at 2.5–300 K was measured using a conventional four-probe method with Ag paste as electrodes. Magnetization (M) measurements were performed with a vibrating sample magnetometer (Quantum Design, San Diego, CA, USA). The electronic structure of the primitive cell was calculated based on linearized/augmented plane wave plus local orbitals (L/APW+lo) method with a density-functional theory level using PBE96 functionals with the WIEN2k code.[7]

The sintered polycrystalline sample, which was dark gray with a metallic luster, was confirmed to be La$_2$Sb by powder XRD measurements that took into consideration the effect of the preferred orientation. Figure 1 shows the crystal structure of La$_2$Sb, which was reported by Calvert et al. in 1970.[5] The characteristics of the structure are described, and then the results of physical property measurements are shown. The structure belongs to the tetragonal *I4/mmm* space group (No. 139) and has three mirror planes. In this structure, the La square net and the rock salt-type LaSb layer are alternately stacked along the c-axis. A similar stacking structure can be seen in PbFCl-type Sc$_2$Sb, which has no mirror plain at z=0.[8] There are two crystallographic sites for La atoms in La$_2$Sb, in which La1 sits on the 4c site with D$_{2h}$-site symmetry, and builds the square net. The La1–La1 distance (3.3 Å), which is much shorter than those of hexagonal-La ($\alpha$-phase) (3.7-3.8 Å),[9] indicates strong metal–metal bonding within the square net (figure 1b). However, both La2 and Sb locate on a 4-fold axis, forming a rock salt-type layer, which is similar to cubic rock salt-type LaSb,[10] where the La–Sb and La–La distances are 3.25, and 4.59 Å, respectively. The Sb anion in the LaSb layer shifts toward the positive square net of La1 along c-axis because of Columbic attraction, and the positive La2 ion moves in the opposite direction, resulting in the distortion of the rock salt-layer. This gives rise to the alternation of the La2–La2 distances (4.12 Å,

4.72 Å) along edge-sharing chain, as shown in figure 1b. The short distance agrees well with La–La distances (3.7-4.1 Å) reported for LaX (X=Cl, or Br), a metal–rich compound.[11] Metal–metal bonds in solids are frequently observed in the compounds of early transition metals, such as Ti, V, Nb, and Mo, where the larger overlap between the d orbitals is possible because of the comparatively small nuclear charge, which allows the large spatial spread of the orbitals.[12] The d–d bonding through an oxygen vacancy in the rock salt-type compounds is also commonly observed in TiO and NbO.[13] However, metal–metal bonds of rare Earth ions are not so common in solids. This is probably because of the chemical instability that originates from the small electronegativity (or small work function).

Figure 2(a) shows resistivity–temperature ($\rho$–T) curve for $La_2Sb$ under an applied magnetic field of 0 Oe. exhibited a metallic behavior, and the value of the curve at 300 K is $2.2\times10^{-4}$ ohm cm. It decreases almost linearly with T for T > 100 K, but switches to a $T^2$-dependence below 30 K. As shown in the inset in Fig. 2(a), a sharp drop in $\rho$ is observed at T = 5.6 K, and the resistivity vanishes at 5.2 K. The critical temperature for zero resistivity shifts downward with increasing H, which suggests that $La_2Sb$ undergoes a superconducting transition at $T_c$=5.2 K. The magnetic susceptibilities ($\chi$) measured in zero-field cooling processes (ZFC) reach −6.0 emu/mol (Figure 2(b)). This value corresponds to a shielding volume fraction of the superconductivity phase of ~130 % (estimated from the $\chi$ value for perfect diamagnetism), substantiating that the bulk superconductive transition takes place at 5.4 K. The M–H curve at 2.5 K, shown in the inset in Fig. 2(b), shows a typical profile for a type-II superconductor with a lower superconducting critical magnetic field ($H_{c1}$) of ~75 Oe.

Figure 3 shows the calculated total density of states (DOS) of $La_2Sb$ together with a partial DOS for the relevant orbitals. The energy zero is taken at $E_F$. The features of the obtained DOS agree basically with the report by Harima at 1993.[14] The bands crossing $E_F$ are in agreement with the metallic electrical properties of La$_2$Sb. The bands near −9 eV are assigned to Sb 5s, and most of the contribution of the Sb 5p orbital is seen in the region of −3.6 – −1 eV, indicating that the valence state of Sb is −3 with a 5s$^2$5p$^6$ closed shell configuration. However, the more positive La is ionized, and the strong peak at +3 eV is ascribed to an unoccupied La 4f state. The La 5d orbitals predominantly contribute to broad bands near E$_F$. In particular, the d$_{xy}$ and d$_{x2-y2}$ orbitals of the La1 square net have a strong d–d σ/π interaction, resulting in the bands being dispersed widely in the xy plane. The position of E$_F$ is located near the peak in the total DOS. Such a situation is often found in the electronic structure of superconductors such as the A15 family.[15] This implies that the high DOS at E$_F$ is the primary origin of the observed superconductivity. Because LaSb has no superconductivity at T$_c$ > 2 K, we tentatively attribute the emergence of superconductivity to the presence of the La square net, which has strong La–La metal bonding.

**Acknowledgments**

This work was supported by the Funding Program for World-Leading Innovative R&D on Science and Technology (FIRST), Japan.

**Notes and references**

**Figure Captions**

**Fig. 1** (a) Crystal structure of La$_2$Sb. The structure is composed of the alternate stacking of La square nets and LaSb layers. (b) Local coordination of these layers and the La–La separation.

**Fig. 2** (a)Temperature (T) dependence of the electrical resistivity ($\rho$) at 0 Oe. The inset shows $\rho$–T curves as a function of the magnetic field. (b) Temperature dependence of the magnetic susceptibility ($\chi$) under conditions of ZFC and FC at 10 Oe. The inset shows the field (H) dependence of the magnetization (M) at 2.5 K.

**Fig. 3** Calculated total density of states of La$_2$Sb with the partial DOS for La1 5d, La2 5d, and Sb 5p.

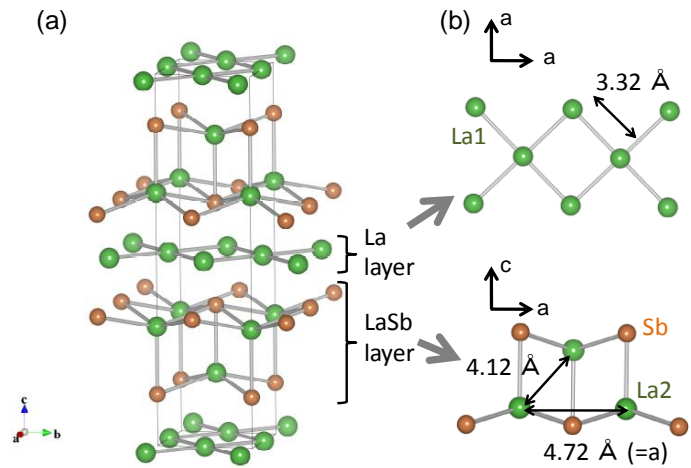

**Figure 1.**

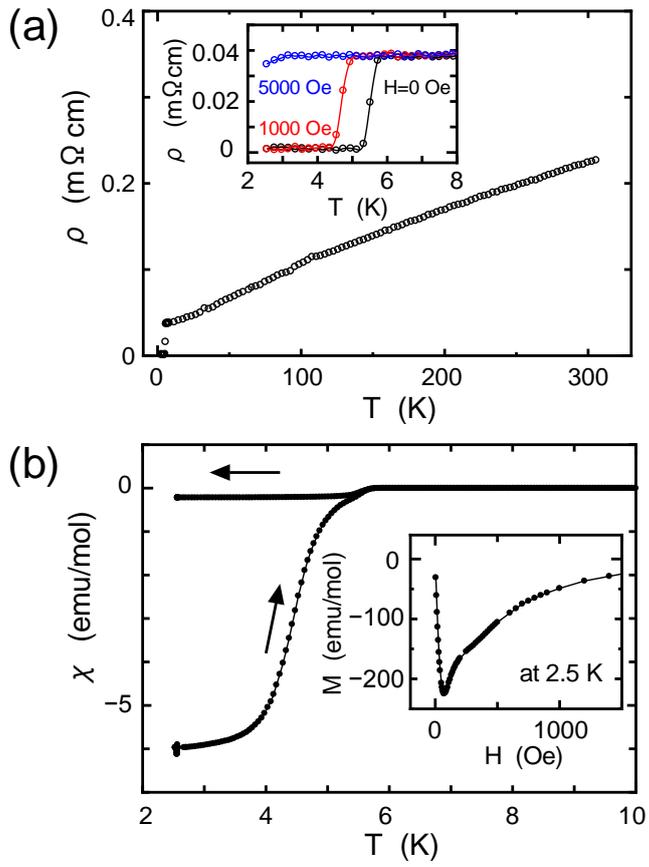

**Figure 2.**

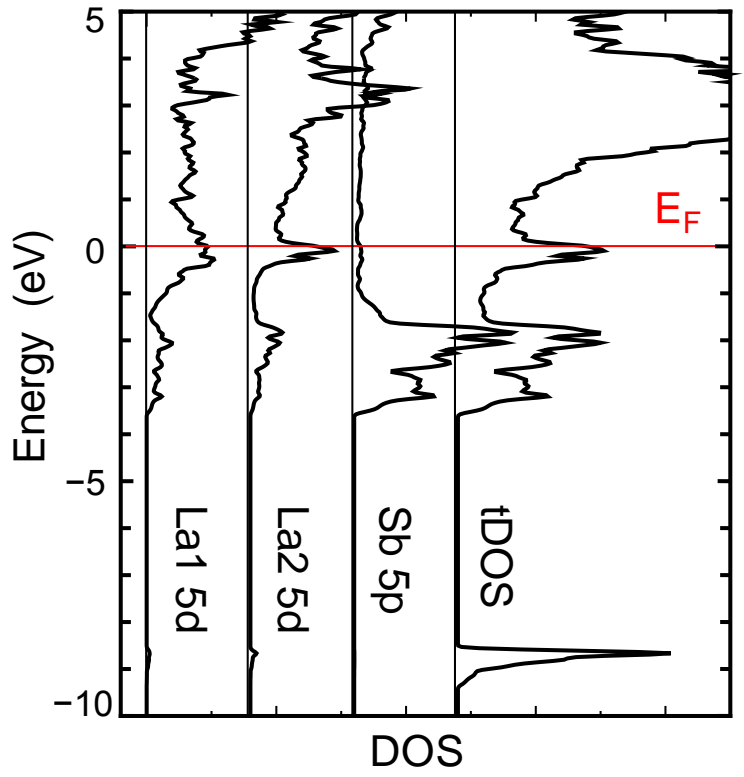

**Figure 3.**